\documentclass[aps,pra,floatfix,twocolumn,superscriptaddress]{revtex4-1}
\usepackage{graphicx}
\usepackage[english]{babel}
\usepackage{amsmath}
\usepackage{amssymb}
\usepackage{tensor}

\newcommand{\vk}{\mathbf{k}}
\newcommand{\vc}{\mathbf{c}}

\newcommand{\be}{\begin{eqnarray}}
\newcommand{\ee}{\end{eqnarray}}
\newcommand{\p}{\partial}

\newcommand{\da}{a^{\dagger}}

\def\ket#1{|#1\rangle}
\def\bra#1{\langle #1 |}
\def\ep#1{\langle #1 \rangle}

\begin{document}

\title{Topological classifications of  quadratic bosonic excitations in closed and open systems with examples}

\author{Yan He}
\affiliation{College of Physics, Sichuan University, Chengdu, Sichuan 610064, China}
\email{heyan$_$ctp@scu.edu.cn}

\author{Chih-Chun Chien}
\affiliation{Department of physics, University of California, Merced, CA 95343, USA.}
\email{cchien5@ucmerced.edu}

\begin{abstract}
The topological classifications of quadratic bosonic systems according to the symmetries of the dynamic matrices from the equations of motion of closed systems and the effective Hamiltonians from the Lindblad equations of open systems are analyzed. While the non-Hermitian dynamic matrix and effective Hamiltonian both lead to a ten-fold way table, the system-reservoir coupling may cause a system with or without coupling to a reservoir to fall into different classes. A 2D Chern insulator is shown to be insensitive to the different classifications. 
In contrast, we present a 1D bosonic Su-Schrieffer-Heeger model with chiral symmetry and a 2D bosonic topological insulator with time-reversal symmetry to show the corresponding open systems may fall into different classes.
\end{abstract}

\maketitle

\section{Introduction}
Studies of topological insulators and topological superconductors according to the underlying symmetries have led to ``periodic tables" that classify various topological systems in different dimensions (see Refs.~\cite{Chiu2016,Stanescu_book} for a review).
For electronic systems, the topological properties are usually analyzed by the band structures because the Fermi-Dirac statistics leads to occupied or unoccupied bands. Since the ground states of bosonic systems correspond to a Bose-Einstein condensate (BEC) of massive bosons or vacuum of massless bosons, topological properties of bosonic systems may be studied in the excited states via the Bogoliubov-de Gennes (BdG) formalism, which has been applied to collective modes of superfluid helium~\cite{Walecka} and atomic BEC~\cite{Ronen06,Pethick_book,FurukawaNJP15,Wang-2020}, photonics~\cite{PeanoNatComm2016}, phononics~\cite{SanavioPRB20}, magnons~\cite{Shindou13,LeinPRB19,KondoPTEP20}, magnetoelastic excitations~\cite{ParkPRB19}, and others~\cite{GurariePRB03,LieuPRB18,Curtis19,Akagi21,Ling21}.

Subtle differences may arise in the classifications according to symmetries as different physical quantities may be chosen to examine the effects of symmetries.
Due to the Bose-Einstein statistics, the Hamiltonian and the equation of motion of bosonic excitations may have different structures~\cite{Zhou20,Xu20,Flynn20,Ashida20}. In contrast, the fermion equation does not show such a complication.
Explicitly, the time evolution of bosonic excitations follows a non-Hermitian dynamic matrix instead of the Hermitian Hamiltonian due to the underlying commutation relations. Moreover, non-Hermitian properties may be present even when the bosonic Hamiltonian is Hermitian~\cite{Yokomizo21}. Through mappings between the dynamics of quadratic bosonic systems and fermionic systems~\cite{Flynn20}, symmetries of single-particle and many-particle systems have been analyzed. Moreover, stability of bosonic systems has been discussed~\cite{Xu20}.

The second-quantized Hamiltonian of a quadratic fermionic (bosonic) system can be diagonalized by a Bogoliubov transformation with the coefficients forming a unitary (para-unitary) matrix in order to satisfy the anti-commutation (commutation) relations~\cite{Walecka}.
For bosons, the energy spectrum can be found equivalently from the diagonalization of the dynamic matrix~\cite{LieuPRB18,KondoPTEP20}. Importantly, the bosonic dynamic matrix transforms under the particle-hole symmetry in the same way as the fermionic Hamiltonian. Therefore, the tenfold-way classification of fermions applies to bosonic excitations according to their dynamic matrix, despite its non-Hermitian property. Interesting topological properties and invariants of non-Hermitian systems have been reviewed in Refs.~\cite{Ghatak19,Ashida20}.

For quantum systems coupled to external reservoirs, the Lindblad equation has been widely used to describe quantum dynamics of open systems~\cite{Haroche_book,Breuer_book,Weiss_book}. There have been many studies of topological properties of open quantum systems~\cite{Diehl2011b,Huang14,SongPRL19,Asorey19,Bandyopadhyay21,McDonald21,Flynn21}. By implementing a method known as the "third quantization" for quadratic fermionic systems~\cite{Prosen08}, Ref.~\cite{Lieu20} provides a classification of the steady states of the Lindblad equation according to the symmetries. Since the third quantization for quadratic bosonic systems has been developed~\cite{Prosen10} as well, here we analyze the topological classification of open quadratic bosonic systems described by the Lindblad equation. The effective Hamiltonian from the Lindblad equation plays the role of the dynamic matrix in the equation of motion of an isolated system.
While non-Hermitian properties will be encountered in the dynamic matrix and effective Hamiltonian of bosons, here we only consider models with real line gaps or models whose complex spectrum can be continuously deformed into some intervals on the real axis. With this constraint, we will not address the full extent of the 38-fold way classification of non-Hermitian models \cite{Kawabata}. In our discussion, the symmetry classes of non-Hermitian models will be reminiscent of the ten-fold way classification of Hermitian fermionic models.

By analyzing concrete examples, we will show that bosonic systems may or may not exhibit different topological properties according to the dynamic matrix, effective Hamiltonian, and their fermionic counterparts. 
We begin with a 2D bosonic Chern insulator that lacks a symmetry, therefore showing similar topological behavior according to different classifications.
To contrast the influence from system-reservoir coupling, we consider a 1D example of a bosonic Su-Schrieffer-Heeger (SSH) model with chiral symmetry. The classifications of the dynamic matrix and effective Hamiltonian agree if the system-reservoir coupling respects chiral symmetry. In that case, the system
shows a quantized winding number with periodic boundary condition and localized edge states with open boundary condition. However, if the system-reservoir coupling breaks chiral symmetry, the effective Hamiltonian may belong to a different class from the one of the dynamic matrix. Our last example is the 2D time-reversal invariant topological insulator described by the four-band Bernevig-Hughes-Zhang (BHZ) model~\cite{BHZ-1}. For the fermionic BHZ model, the topological properties have been demonstrated in semiconductor quantum wells~\cite{BHZ-2}. We construct a bosonic version similar to that studied in Ref.~\cite{Wang-2020} but consider both closed and open systems. A comparison shows that in the presence of Lindblad operators breaking time-reversal symmetry, the effective Hamiltonian from the Lindblad equation falls into a different class compared to the dynamic matrix of the bosonic BHZ model.

The rest of the paper is organized as follows. Sec.~\ref{sec:classification} summarizes the classifications of quadratic bosonic systems in closed and open settings via symmetries and dimensions. The ten-fold way table according to the dynamic matrix from the equation of motion or the effective Hamiltonian from the Lindblad equation are explained. To contrast the similarities and differences, Sec.~\ref{sec:example} shows a 2D bosonic Chern insulator that does not differentiate the various classifications. Meanwhile, a 1D bosonic SSH model and a 2D bosonic BHZ model are shown to exhibit different topological behavior due to system-reservoir couplings that break chiral or time-reversal symmetry in the effective Hamiltonian. Sec.~\ref{sec:conclusion} concludes our work. In the Appendix, we summarize some details and subtleties.

\section{Symmetries and Classifications of Quadratic Bosonic systems} \label{sec:classification}
The generic Hermitian Hamiltonian of a quadratic bosonic system may be written as
\be
\mathcal{H}=\sum_{a,b}\psi_a^{\dag}H_{ab}\psi_b,\qquad
H=\left(\begin{array}{cc}
         A & B \\
          B^* & A^*
\end{array}\right).
\label{BdG}
\ee
Here we define $\psi=(a_1,\cdots,a_n,a_1^\dag,\cdots,a_n^\dag)^T$. The matrices $A$ and $B$ satisfy $A^{\dag}=A$ and $B^T=B$. There has been an attempt \cite{Zhou20} to classify the topology of bosonic system according to the Hamiltonian, but the connection to the conventional classification should be built on the dynamic matrix~\cite{GurariePRB03}. We summarize the classification in the next subsection and then generalize the classification to open systems described by the Lindblad equation.

\subsection{Classification according to dynamic matrix of closed systems}
The dynamic matrix comes from the equation of motion of a quadratic bosonic system with the standard commutation relation: ($\hbar\equiv 1$)
\be
i\frac{d\psi_i}{dt}=[\psi_i, H]=\sum_j\Big(\tau_3 H\Big)_{ij}\psi_j.
\ee
Note that the boson operators satisfy the following canonical commutation relations:
\be
[\psi_i,\psi_j^\dag]=(\tau_3)_{ij},\quad
\psi=\Big(a_1,\cdots,a_n,\da_1,\cdots,\da_n\Big)^T.
\ee
Here we have introduced $\tau_i=\sigma_i\otimes I$, which applies to the "Nambu space" of bosons \cite{Wang-2020}.
The dynamic matrix is given by
\be
H_{\textrm{dyn}}=\tau_3H=\left(\begin{array}{cc}
         A & B\\
         -B^* & -A^*
\end{array}\right).
\ee
For fermionic systems, the dynamic matrix coincides with the Hamiltonian due to the fermionic anti-commutation relations.
In the following, we will show that the dynamic matrix $H_{\textrm{dyn}}$ of bosons may behave differently under symmetry transformations when compared to the Hermitian bosonic Hamiltonian $H$.

We apply the ideas of Refs.~\cite{Kawabata,Lieu20} and use the three types of discrete symmetries to classify the dynamic matrix $H_{\textrm{dyn}}$ of a quadratic boson system.
We first discuss the time-reversal symmetry, under which the bosonic operators transform as
\be
\mathcal{T} a_i\mathcal{T}^{\dag}=\sum_j (T)_{ji}a_j,\quad
\mathcal{T} a_i^\dag\mathcal{T}^{\dag}=\sum_j (T)^*_{ji}a_j^\dag.
\ee
Here $\mathcal{T}$ is the time-reversal operator in the second-quantization form. In the first-quantization form, the time-reversal symmetry of the Hamiltonian gives the condition
\be
\left(\begin{array}{cc}
         T & 0\\
         0 & T^*
\end{array}\right)^\dag\left(\begin{array}{cc}
         A^* & B^*\\
         B & A
\end{array}\right)\left(\begin{array}{cc}
         T & 0\\
         0 & T^*
\end{array}\right)=\left(\begin{array}{cc}
         A & B\\
         B^* & A^*
\end{array}\right),
\ee
which in turn determines the transformations of the matrices $A$ and $B$ as
\be
T^\dag A^* T=A,\quad T^\dag B^* T^*=B.
\ee
Then, the time-reversal of the dynamic matrix $H_{\textrm{dyn}}$ is
\be
\left(\begin{array}{cc}
         T & 0\\
         0 & T^*
\end{array}\right)^\dag\left(\begin{array}{cc}
         A^* & B^*\\
         -B & -A
\end{array}\right)\left(\begin{array}{cc}
         T & 0\\
         0 & T^*
\end{array}\right)=\left(\begin{array}{cc}
         A & B\\
         -B^* & -A^*
\end{array}\right). \nonumber \\
\ee
Although $H_{\textrm{dyn}}$ of a bosonic system may be non-Hermitian, it satisfies the time-reversal symmetry condition $T_1^\dag H_{\textrm{dyn}}^*T_1=H_{\textrm{dyn}}$ with $T_1=\textrm{diag}(T,T^*)$. Therefore, the time-reversal symmetry of the dynamic matrix $H_{\textrm{dyn}}$ of the bosonic system behaves just like a Hermitian fermionic Hamiltonian.

\begin{table}
  \centering
  \begin{tabular}{c|ccc|ccccccc}
  \hline
  Class & T & C & S & $d=0$ & $d=1$ & $d=2$ & $d=3$ & $d=4$ & $d=5$ & $d=6$ \\
  \hline
  A     & 0 & 0 & 0 & $\mathbb{Z}$ & 0 & $\mathbb{Z}$ & 0 & $\mathbb{Z}$ & 0 & $\mathbb{Z}$ \\
  AIII  & 0 & 0 & 1 & 0 & $\mathbb{Z}$ & 0 & $\mathbb{Z}$ & 0 & $\mathbb{Z}$ & 0 \\
  \hline
  AI    & + & 0 & 0 & $\mathbb{Z}$ & 0 & 0 & 0 & $\mathbb{Z}$ & 0 & $\mathbb{Z}_2$ \\
  BDI   & + & + & 1 & $\mathbb{Z}_2$ & $\mathbb{Z}$ & 0 & 0 & 0 & $\mathbb{Z}$ & 0 \\
  D     & 0 & + & 0 & $\mathbb{Z}_2$ & $\mathbb{Z}_2$ & $\mathbb{Z}$ & 0 & 0 & 0 & $\mathbb{Z}$ \\
  DIII  & - & + & 1 & 0 & $\mathbb{Z}_2$ & $\mathbb{Z}_2$ & $\mathbb{Z}$ & 0 & 0 & 0 \\
  AII   & - & 0 & 0 & $\mathbb{Z}$ & 0 & $\mathbb{Z}_2$ & $\mathbb{Z}_2$ & $\mathbb{Z}$ & 0 & 0 \\
  CII   & - & - & 1 & 0 & $\mathbb{Z}$ & 0 & $\mathbb{Z}_2$ & $\mathbb{Z}_2$ & $\mathbb{Z}$ & 0 \\
  C     & 0 & - & 0 & 0 & 0 & $\mathbb{Z}$ & 0 & $\mathbb{Z}_2$ & $\mathbb{Z}_2$ & $\mathbb{Z}$ \\
  CI    & + & - & 1 & 0 & 0 & 0 & $\mathbb{Z}$ & 0 & $\mathbb{Z}_2$ & $\mathbb{Z}_2$ \\
  \hline
\end{tabular}
  \caption{Classification of symmetries and topology of quadratic bosonic systems according to their dynamic matrices. The first column shows the Cartan labels of the ten classes. The next three columns show the properties of these classes under time-reversal, particle-hole and chiral symmetries. Here $0$ or $1$ represents the absence or presence of the symmetry. The $+$ and $-$ signs correspond to $TT^*=\pm1$ and $CC^*=\pm1$. The last seven columns show the possible topological phases of the ten classes in various dimensions. The classification according to the effective Hamiltonian produces the same table, but the system-reservoir coupling may change which class a system belongs to.}
  \label{tenfold-1}
\end{table}

Next, we turn to the particle-hole symmetry of the dynamic matrix. Explicitly, the transformation matrix for the particle-hole symmetry may be taken as $C=\tau_1$, leading to
\be
\tau_1 H_{\textrm{dyn}}^*\tau_1=-H_{\textrm{dyn}}.
\ee
Therefore, we arrived at the same particle-hole symmetry condition as a quadratic fermionic model. The composition of time-reversal and particle-hole symmetries gives rise to chiral symmetry. The condition $(CT)H_{\textrm{dyn}}(CT)^{\dag}=-H_{\textrm{dyn}}$ behaves in the same way as the fermionic case does. In summary, for the non-Hermitian dynamic matrix $H_{\textrm{dyn}}$ of bosons, the three discrete symmetry conditions are
\be
&&\textrm{Time-reversal}:~~ T H_{\textrm{dyn}}^* T^{\dag}=H_{\textrm{dyn}},~~ T T^*=\pm1,\\
&&\textrm{Particle-hole}:~~ C H_{\textrm{dyn}}^*C^{\dag}=-H_{\textrm{dyn}},~~ C C^*=\pm1,\\
&&\textrm{Chiral or sublattice}:~ S H_{\textrm{dyn}} S^{\dag}=-H_{\textrm{dyn}},~ S^2=1.
\ee
Since the symmetry conditions are the same as the quadratic fermionic case, the ten-fold way classifications of quadratic fermionic systems \cite{Chiu2016} can be applied to the dynamic matrix of quadratic boson systems. The classification is summarized in Table \ref{tenfold-1}. We remark that 
particle-hole symmetry is only significant for systems with particle-hole mixing, which arises in systems with pairing effects. Therefore, models without particle-hole mixing do not discern the difference because particle-hole symmetry has no effect on them.

\subsection{Classification according to effective Hamiltonian of open systems}
We now turn to open quantum systems of bosons. The Lindblad quantum master equation for the time evolution of the reduced density matrix $\rho$ under the influence of an environment with the Markovian approximation has the expression~\cite{Haroche_book,Weiss_book,Breuer_book} ($\hbar=1$)
\be\label{Lindblad}
i\frac{d\rho}{dt}=\mathcal{L}(\rho)=[\mathcal{H},\rho]+i\sum_{\mu}\Big(2L_{\mu}\rho L_{\mu}^{\dag}-\{L_{\mu}^{\dag}L_{\mu},\rho\}\Big).
\ee
Here $L_\mu$ are the Lindblad operators modeling the environmental effects on the system.
Assuming a quadratic bosonic system with $n$ indices, the generic Hamiltonian is given by
\be
\mathcal{H}=\sum_{i,j}\Big(\da_i A_{ij} a_j+a_i A^*_{ij} \da_j+a_i B_{ij} a_j+\da_i B_{ij}^* \da_j\Big).
\label{eq-H}
\ee
Here the  $A$ and $B$ satisfy $A^{\dag}=A$ and $B^T=B$. To obtain an exact expression for classifying open bosonic systems, we consider linear Lindblad operators similar to those used in Ref.~\cite{Prosen10}:
\be
L_{\mu}=\sum_j \Big(l_{\mu j}a_j+k_{\mu j}\da_j\Big).\label{eq-L}
\ee
We will focus on open systems that allow a steady-state solution of the Lindblad equation in the long-time limit.

Following  Ref.~\cite{Prosen10} and Appendix~\ref{App:3rdQ}, the Liouvillean can be rewritten as
\be
\mathcal{L}=\left(
              \begin{array}{cc}
                \vc' & \vc
              \end{array}
            \right)\left(
                     \begin{array}{cc}
                       -X^T & Y \\
                       0 & -X
                     \end{array}
                   \right)\left(
                            \begin{array}{c}
                              \vc \\
                              \vc'
                            \end{array}
                          \right).
                          \label{L-op}
\ee
Here $\vc=(c_{0,1}\cdots,c_{0,n},c_{1,1}\cdots,c_{1,n})^{T}$ and $\vc'=(c'_{0,1}\cdots,c'_{0,n},c'_{1,1}\cdots,c'_{1,n})^{T}$ are the transformed bosonic operators. We define the following $2n\times 2n$ matrices
\be
&&X=\left(\begin{array}{cc}
         -A^* & B \\
         -B^* & A
\end{array}\right)+\frac{i}{2}\left(\begin{array}{cc}
         M-N^* & L^T-L \\
         L^{\dag}-L^* & M^*-N
\end{array}\right),\\
&&Y=\frac{i}{2}\left(\begin{array}{cc}
         -2iB^*-L^*-L^{\dag} & 2N \\
         2N^T & 2iB-L-L^T
\end{array}\right).
\ee
Here we define three $n\times n$ matrices
\be
M_{ij}=\sum_{\mu} l_{\mu i}l^*_{\mu j},~ N_{ij}=\sum_{\mu} k_{\mu i}k^*_{\mu j},~
L_{ij}=\sum_{\mu} l_{\mu i}k^*_{\mu j}. \nonumber \\
\ee
Ref.~\cite{Lieu20} proposed that the effective-Hamiltonian $X$ in the open quantum system may play the role of the Hamiltonian in the corresponding isolated system. In general, $X$ is a non-Hermitian matrix. Its $2n$ complex eigenvalues $\lambda_1,\cdots,\lambda_{2n}$ are the counterpart of the eigen-energies of the  Hamiltonian of an isolated system.

Following the idea of Ref.~\cite{Lieu20}, we can also use the three types of discrete symmetries to classify the effective Hamiltonian $X$ from the Lindblad equation modeling a bosonic system influenced by the environment. For reasons that will be explained shortly, we decompose the effective Hamiltonian as
\be
X=X_0+X_1=\left(\begin{array}{cc}
         -A^* & B \\
         -B^* & A
\end{array}\right)+\frac{i}{2}\left(\begin{array}{cc}
         M-N^* & L^T-L \\
         L^{\dag}-L^* & M^*-N
\end{array}\right). \nonumber \\
\ee
The Hamiltonian part gives $X_0$ and the jump operator part gives $X_1$.
In what follows, we will first discuss the time-reversal symmetry. The key step is that the linear Lindblad operator satisfies
\be
\mathcal{T}_S L_{\mu}^*\mathcal{T}_S^{\dag}=L_{\mu}.
\ee
Here $\mathcal{T}_S$ is the time-reversal operator applying to the system only. Recall that the boson operators transform under time reversal as
\be
\mathcal{T}_S c_i\mathcal{T}_S^{\dag}=\sum_j (T_S)^*_{ji}c_j,\quad
\mathcal{T}_S c_i^\dag\mathcal{T}_S^{\dag}=\sum_j (T_S)_{ji}c_j^\dag.
\ee
This leads to the following relations
\be
\sum_i (T_S)^*_{ij}l^*_{\mu,j}=k_{\mu,i},\quad
\sum_i (T_S)_{ij}k^*_{\mu,j}=l_{\mu,i},
\ee
which in turn determine the transformations of the matrices $M$, $N$ and $L$ as
\be
T_S M T_S^{\dag}=N^T,\quad T_S^* N T_S^T=M^T,\quad C_S L C_S^T=L^T.
\ee
Note that both $M$ and $N$ are Hermitian matrices. Then the time-reversal of the jump operator part $X_1$ is
\be
&&T X_1 T^\dag=-X_1,\\
&&T=\left(\begin{array}{cc}
         T_S & 0\\
         0 & T^*_S
\end{array}\right),~
X_1=\frac{i}{2}\left(\begin{array}{cc}
         M-N^* & L^T-L \\
         L^{\dag}-L^* & M^*-N
\end{array}\right).
\ee
The Hamiltonian part $X_0$ is similar to the dynamic matrix $H_{\textrm{dyn}}$, which satisfies the ordinary time-reversal symmetry condition $TX_0^*T^{\dag}=X_0$. If we assume the coefficients $l_{\mu,i}$ and $k_{\mu,i}$ in the jump operators are all real numbers, it follows that the jump operator part $X_1$ also satisfies $T X_1^*T^{\dag}=X_1$. Combining these results, we arrived at $T X^*T^{\dag}=X$. Therefore, in an open bosonic system described by the Lindblad equation, the time-reversal symmetry condition is applied to the complex conjugate of the effective Hamiltonian $X$, similar to the case of a fermionic Hamiltonian. However, this is different from the fermionic effective Hamiltonian of the Lindblad equation, where the time-reversal condition is applied to the transpose of the effective Hamiltonian \cite{Lieu20}.

Now we turn to the particle-hole symmetry. Since $X$ is from the dynamic equation of $\rho$, we expect its symmetry transformations to be similar to those of $H_{\textrm{dyn}}$ from the equation of motion rather than the Hermitian bosonic Hamiltonian $H$. Explicitly, the transformation matrix for the particle-hole symmetry can be taken as $C=\tau_1$, leading to
\be
\tau_1 X_0^*\tau_1=-X_0,\quad \tau_1 X_1^*\tau_1=-X_1.
\ee
Therefore, we arrived at the particle-hole symmetry condition $\tau_1 X^*\tau_1=-X$, which is the same as the quadratic fermionic model. The composition of the time-reversal and particle symmetries gives rise to the chiral symmetry condition $(CT)X(CT)^{\dag}=-X$, which is also the same as the fermionic case. In summary, for the non-Hermitian effective Hamiltonian $X$, the three discrete symmetry conditions are
\be
&&\textrm{Time-reversal}:~ T X^* T^{\dag}=X,~ T T^*=\pm1,\\
&&\textrm{Particle-hole}:~ C X^*C^{\dag}=-X,~ C C^*=\pm1,\\
&&\textrm{Chiral or sublattice}:~ S X S^{\dag}=-X,~ S^2=1.
\ee
Since the symmetry conditions are the same as the quadratic fermionic case, the ten-fold way classifications of quadratic fermionic systems \cite{Chiu2016} can also be applied to the effective Hamiltonian $X$ of quadratic boson systems described by the Lindblad equation with linear Lindblad operators. Therefore, the classification produces the same table as Table \ref{tenfold-1}. However, the system-reservoir coupling in the Lindblad equation may lead to an explicit non-Hermitian effective Hamiltonian, causing subtle differences between the same model with and without the coupling to a reservoir.

Before demonstrating the subtle differences in the different classifying schemes by examples, we comment on the physical relevance of the schemes. For a closed bosonic system, the spectrum is determined by the dynamic matrix from the equation of motion or BdG diagonalization, so the classification follows Table~\ref{tenfold-1}. The quasi-particle states may be used to determine the properties in this case. 
Moreover, topological properties usually involve dynamic changes of the system~\cite{Chiu2016,Stanescu_book}, the classification according to the dynamic matrix prevails if the system is isolated. 
If the bosonic system is influenced by an environment and describable by the Lindblad formalism, the classification also follows Table~\ref{tenfold-1} when one focuses on the effective behavior of the system. The properties may be extracted from the density matrix and its spectrum from the Lindblad equation, which incorporates the effects from the environment.

\section{Examples}\label{sec:example}
\subsection{2D Chern insulator}
\begin{figure}
\centering
\includegraphics[width=\columnwidth]{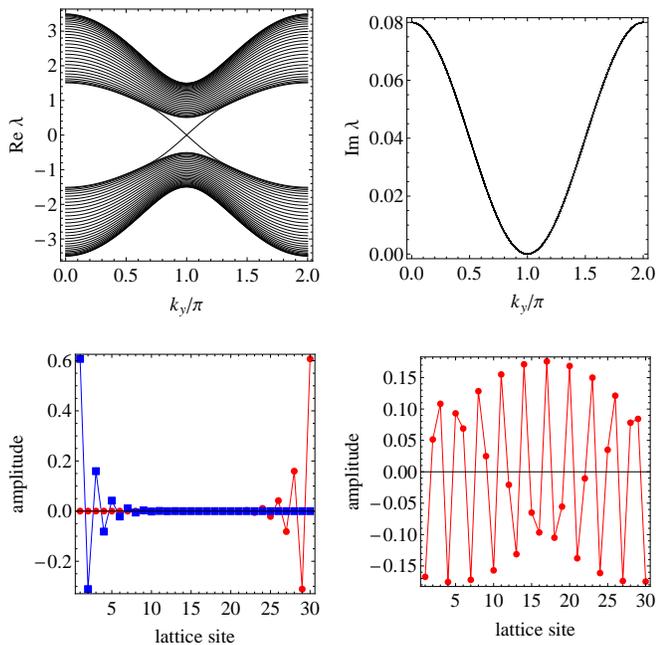}
\caption{The top row shows the real (left) and imaginary (right) parts of the spectrum of the effective Hamiltonian of the 2D bosonic Chern insulator as a function of $k_y$ for $m=1.5$ and $\gamma=0.2$. The bottom row shows the wave functions of the two edge states (left) and a typical bulk state (right) along the direction with open boundary condition.}
\label{fig-CI}
\end{figure}

We first give an example that does not differentiate the classifications of bosons in the presence of system-reservoir coupling. The example is a 2D model in class A without any symmetry, which can be thought of as the bosonic counterpart of the fermionic Chern insulator. In real space, the Hamiltonian is given by
\be
\mathcal{H}&=&\sum_{n}\Big(\da_{n}\frac{\sigma_3-i\sigma_1}{2}a_{n+\hat{x}}
+\da_{n}\frac{\sigma_3-i\sigma_2}{2}a_{n+\hat{y}}+h.c.\Big) \nonumber \\
& &+\sum_{n} m \da_{n,s} \sigma_3 a_{n}.\label{CI}
\ee
Here $n=(n_x,n_y)$ labels the lattice site on a square lattice. With two orbitals labelled by $1,2$ on each site, the boson operator is defined as $a_{n}=(a_{n,1}, a_{n,2})^T$. We also define $\hat{x}$ and $\hat{y}$ as the unit vectors along the $x$ and $y$ directions, respectively.  There have been studies of the bosonic Chern insulator~\cite{Shindou13,PeanoNatComm2016}, but here we analyze the model in both closed- and open- system settings.
We remark that there is practically no difference between the bosonic and fermionic Chern insulators because the Chern insulator does not have any of the three classifying symmetries. Hence, the complication from the commutation and anticommutation relations does not play a significant role here. As a consequence, there is no difference between the classifications for the 2D bosonic and fermionic Chern insulators. Nevertheless, here we present the topological properties of the 2D bosonic Chern insulator described by the Lindblad equation.

The Lindblad operators we considered here have the form
\be
L_{n,i}=\gamma(a_{n,i}+a_{n+\hat{y},i})\quad \textrm{for}\quad i=1,2.
\ee
Transforming the expression to momentum space, the effective Hamiltonian is given by
\be
X=\left(\begin{array}{cc}
         -H^*+i M/2 & 0 \\
         0 & H+i M^*/2
\end{array}\right).
\ee
The matrices $H$ and $M$ are defined by
\be
&&H=\sum_{j=1}^3R_j\sigma_j,\quad M=2\gamma^2(1+\cos k_y) I_0, \\
&&R_1=\sin k_x,~ R_2=\sin k_y,~ R_3=m+\cos k_x+\cos k_y. \nonumber \\
\ee
Here $\sigma_j$ with $j=1,2,3$ are the Pauli matrices and $I_0$ is the $2\times2$ identity matrix. The energy spectrum of the effective Hamiltonian is given by
\be
\lambda=\pm\sqrt{R_1^2+R_2^2+R_3^2}+i\gamma^2(1+\cos k_y).
\ee
The topology of this model may be characterized by the Chern number from the effective Hamiltonian. For the lower band, the Chern number can be related to the 2D winding number as
\be
C=-\frac1{4\pi}\int d^2k\epsilon_{ijk}\hat{R}_i\cdot(\p_x\hat{R}_j\times\p_y\hat{R}_k).
\label{wind}
\ee
For the 2D bosonic model discussed here, the Chern number takes the following values:
\be
C=\left\{
    \begin{array}{ll}
      1, & 0<m<2,\\
      -1, & -2<m<0, \\
      0, & |m|>2.
    \end{array}
  \right.
\label{QWZ}
\ee

For non-Hermitian models, one may introduce more types of topological invariants, such as the vorticity in the energy spectrum~\cite{Ghatak19}, defined as
\be
v_n=\frac{1}{2\pi}\oint\nabla_{\vk}\arg\Big[\lambda_n(\vk)\Big]\cdot d\vk.
\ee
The vorticity counts the number of exceptional points inside the contour, and it is associated with the complex point gap~\cite{ShenPRL18,Kawabata,McClarty19,Bergholtz21}. For the model studied here, however, the contribution from the system-reservoir coupling to the effective Hamiltonian of the Lindblad equation, $M$, is proportional to the identity matrix due to the choice of the Lindblad operators. Thus, there is no exceptional point, and the vorticity is trivial. For the discussion of Table~\ref{tenfold-1}, it is more suitable to use the Chern number to characterize the topology. Nevertheless, one can see that this analysis offers another example that different topological classifications can be characterized by different physical quantities.

The Chern number is defined for systems with periodic boundary condition. We further consider the model on a cylinder-shape lattice. Explicitly, we impose open boundary condition along the $x$ axis and periodic boundary condition along the $y$ axis. The resulting Hamiltonian and effective Hamiltonian are still a function of $k_y$. In Figure \ref{fig-CI}, we plot the eigenvalue spectrum of the effective Hamiltonian $H_{\textrm{eff}}$ as a function of $k_y$. We have assumed $m=1.5$, which corresponds to $C=1$. In the upper left panel, one can see that there are two chiral edge states connecting the two bands. The edge states are localized at the two open ends of the cylinder. Moreover, there is no non-Hermitian skin effects because the non-Hermitian terms are diagonal. The typical profiles of the edge and bulk states are shown in the bottom of Figure \ref{fig-CI}. We remark that the system with $C=0$ in the same cylinder geometry shows no chiral edge state.

\subsection{1D bosonic Su-Schrieffer-Heeger (SSH) model}
Our first example to show the system-reservoir coupling may affect the classification is a 1D bosonic SSH model.
Its Hamiltonian in real space is given by
\be
H&=&\sum_i\Big[t_1\da_{1,i} a_{2,i}+t_2\da_{2,i}a_{1,i+1}+\text{H.c.}\Big]
\ee
Here $a_{1,i}$ and $a_{2,i}$ are the annihilation operators of bosons on sublattices $1$ and $2$, respectively, in unit cell $i$. $t_{1,2}$ denote the intra and inter-cell hopping coefficients, respectively.

\begin{figure}
\centering
\includegraphics[width=\columnwidth]{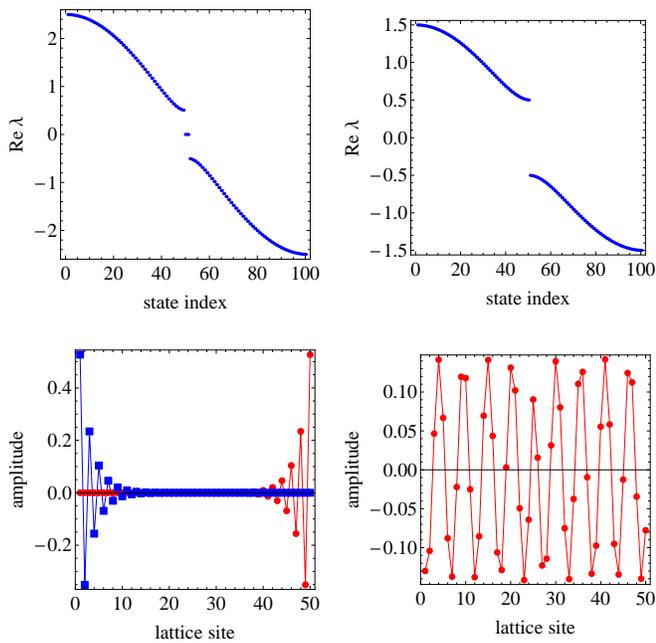}
\caption{(Top row) Spectrum of the effective Hamiltonian of the 1D bosonic SSH model with open boundary condition for $t_2/t_1=1.5$ (left) and $t_2/t_1=0.5$ (right). Here $\gamma/t_1=0.2$. (Bottom row) Wave functions of the edge states (left) and a bulk state (right) on a finite lattice of $50$ sites.}
\label{fig-K}
\end{figure}

For the 1D bosonic SSH model, the classifications according to the dynamic matrix and the effective Hamiltonian are the same if
the Lindblad operators do not break the chiral symmetry. For example, we choose the Lindblad operators as
\be
L_j=\gamma(a_{1,j}+a_{2,j}).
\ee
In momentum space, we find that the effective Hamiltonian from the Lindblad equation is given by
\be
X&=&\left(\begin{array}{cc}
         0 & t_1+t_2 e^{i k}\ \\
         t_1+t_2 e^{-i k} & 0
\end{array}\right)
+\frac{i}{2}\gamma^2\left(\begin{array}{cc}
         1 & 1 \\
         1 & 1.
\end{array}\right).
\ee
The spectrum of the above model is given by
\be
&&\lambda=\pm\lambda_0 +\frac i2\gamma^2.\\
&&\lambda_0=\sqrt{(t'_1+t_2\cos k)^2+t_2^2\sin^2k}.\nonumber
\ee
Here $t'_1=t_1+i\gamma^2/2$, which is also complex. The topology of this model is characterised by the winding number~\cite{Ghatak19,Ashida20}. To compute the winding number, we first introduce a  flattened Hamiltonian as
\be
&&Q=\ket{u^R_1}\bra{u^L_1}-\ket{u^R_2}\bra{u^L_2}
=\left(
   \begin{array}{cc}
     0 & q_1 \\
     q_2 & 0 \\
   \end{array}
 \right),\\
&&q_{1,2}=(t'_1+t_2 e^{\pm i k})/\lambda_0. \nonumber
\ee
Here $\ket{u^R_i}$ and $\ket{u^L_i}$ for $i=1,2$ are the right and left eigenstates of the non-Hermitian effective Hamiltonian $X$. Then, the winding number is defined as 
\be
W=\frac{1}{2\pi}\int_0^{2\pi}d k\, q_1^{-1}\frac{d q_1}{d k}.
\ee
 For the above model, we find that
\be
W=\left\{
    \begin{array}{ll}
      1, & t_1<t_2,\\
      0, & t_1>t_2.
    \end{array}
  \right.
\ee
Next, we will study the system with open boundary condition.

For the 1D bosonic SSH model with open boundary condition, we plot the real part of the spectrum of the effective Hamiltonian in the top row of Figure \ref{fig-K} for $t_2/t_1=1.5$, $t_2/t_1=0.5$ and $\gamma/t_1=0.2$, respectively. Both spectra are gapped, but there are two in-gap states connecting the two bands in the upper-left panel. In contrast, there is no in-gap state in the upper-right panel. Those in-gap states are localized edge states, as shown in the lower-left panel of Figure \ref{fig-K}. To contrast the wavefunction profiles, we also plot the wavefunction of a typical bulk state showing no localization. The in-gap edge states emerge in the regime where the winding number takes the nontrivial value $W=1$. When $W=0$, the system is topologically trivial and there is no localized edge state.

After subtracting a constant matrix, it can be verified that $X'=X-\frac i2\gamma^2$ respects chiral symmetry because
\be
\sigma_3 X'(k) \sigma_3=-X'(k).
\ee
Here $\sigma_i$ are the Pauli matrices applying to the sublattice space. Therefore, the effective Hamiltonian of the 1D bosonic SSH model belongs to class AIII of Table~\ref{tenfold-1}, which supports a $\mathbb{Z}$ classification in 1D. The bulk-edge correspondence is valid in the above example, since the winding number equals to the number zero modes located at one edge. The winding number is also robust against any perturbations that respect chiral symmetry.

In the above example, the effective Hamiltonian and the dynamic matrix belong to the same symmetry class. One can obtain a different effective Hamiltonian by using a different set of Lindblad operators. To demonstrate the possibility of having the dynamic matrix and effective Hamiltonian in different classes, we consider the Lindblad operators
\be
L_j=\gamma a_{2,j}.
\ee
Note that for a time-reversal invariant system, we should impose the condition that the coefficients of the Lindblad operators are real numbers. However, the 1D  bosonic SSH model does not have time-reversal symmetry. Therefore, we are allowed to use complex numbers in the example. In momentum space, the effective Hamiltonian is given by
\be
X=H_{\textrm{dyn}}+\frac{i}{2}\gamma^2\left(\begin{array}{cc}
         0 & 0 \\
         0 & 1.
\end{array}\right).
\ee
It can be shown that $\sigma_3 X(k) \sigma_3\neq-X(k)$, so the Lindblad operators break the chiral symmetry in $X$ even though $H_{\textrm{dyn}}$ respects the symmetry. As a consequence, $X$ belongs to the class A, which is topologically trivial in 1D and different from the classification according to $H_{dyn}$.

We would like to mention that effects of spontaneously symmetry breaking on topological systems have been studied in Ref.~\cite{Raj21}.
Spontaneous symmetry breaking means the ground state of a system does not respect a symmetry of the Hamiltonian. In our classifications of bosonic systems, we analyze the symmetries of the dynamic matrix or its counterpart in open quantum systems. Therefore, the classifications do not take into consideration the broken symmetry in the ground state. 
We remark that there exist examples~\cite{Raj21} where the topology remains the same even though the ground states have different symmetries. Our discussions do not exclude those possibilities.

\subsection{2D time-reversal invariant topological insulator}
As another example for demonstrating the different behavior in closed and open systems, we consider the 2D time-reversal invariant topological insulator described by the four-band Bernevig-Hughes-Zhang (BHZ) model~\cite{BHZ-1}. The bosonic BHZ model as an isolated system has been discussed in Ref.~\cite{Wang-2020}, and here we will further generalized it to include the environmental coupling. For experimental realization, we consider spin-1 bosonic atoms with the $S_z=0$ component projected out~\cite{SpinorBECRMP}. The corresponding bosonic fields are $\psi=(\psi_{1\uparrow},\psi_{2\uparrow},\psi_{1\downarrow},\psi_{2\downarrow})^T$, where the index $i=1,2$ labels the two orbitals and the up or down arrow labels $S_z=\pm 1$. The Hamiltonian of the BHZ model is given by
\be
H_{BHZ}=\left(
    \begin{array}{cc}
      H_0(\vk) &  H_1 \\
      H_1^{\dag} & H_0^*(-\vk)
    \end{array}
  \right).
  \label{eq:BHZ}
\ee
The $H_0$ term is a two-band Chern insulator considered in the previous subsection with the following expression
\be
H_0=\sin k_x \sigma_1+\sin k_y \sigma_2+(m+\cos k_x+\cos k_y)\sigma_3.
\label{QWZ}
\ee
Here $\sigma_i$ for $i=1,2,3$ are the Pauli matrices. The $H_1$ term is given by
\be
H_1=\left(
    \begin{array}{cc}
      0 &  \gamma \\
      -\gamma & 0
    \end{array}
  \right),
\ee
which couples the two spin components. It breaks the $S_z$ conservation and the inversion symmetry, thereby making the model more generic. 
Since the $S_z=\pm 1$ states correspond to a time-reversal pair, the time-reversal operator is giving by $U_TK$ with $U_T=i\sigma_2$ and $K$ denoting the complex conjugation. It follows that
\be
U_T^\dag H^*(\vk) U_T=H(-\vk).
\ee
Thus, the bosonic BHZ model belongs to the AII class with a $\mathcal{Z}_2$ topological index protected by the time-reversal symmetry similar to its fermionic counterpart.

For the BHZ model, the topological index is the Fu-Kane invariant~\cite{Fu-Z2} that can be constructed from the matrix $W$ with the following elements
\be\label{eq:W}
W_{ij}(\vk)=\ep{u_i(-\vk)|U_T K|u_j(\vk)}.
\ee
Here $\ket{u_i(\vk)}$ denotes the eigenstate in momentum space and the indices $i,j$ run through all the occupied bands. $W$ is an anti-symmetric matrix at the time reversal invariant momentum points $\vk_1=0$, $\vk_2=(\pm\pi,0)$, $\vk_3=(0,\pm\pi)$, and $\vk_4=(\pm\pi,\pm\pi)$. The Fu-Kane index can be obtained from the expression
\be
I_{FK}=\prod_{i=1}^4\frac{\textrm{Pf}\,[W(\vk_i)]}{\sqrt{\det W(\vk_i)}}.
\ee
Here the product runs through all the time reversal invariant momentum points, and ``Pf'' denotes the Pfaffian. For the BHZ model with small $\gamma$, the Fu-Kane invariant takes the values
\be
I_{FK}=\left\{
         \begin{array}{ll}
           -1, & |m|<2, \\
           1, & |m|>2.
         \end{array}
       \right.
\ee

\begin{figure}
\centering
\includegraphics[width=\columnwidth]{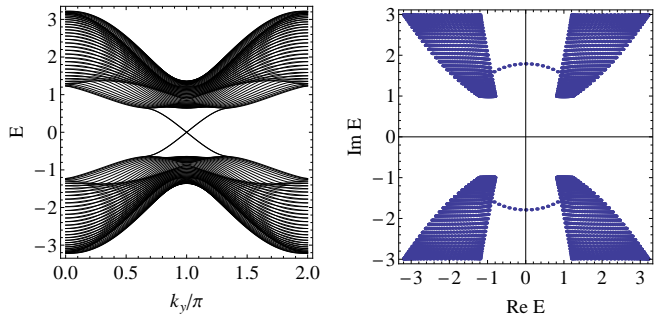}
\caption{(Left) Band structure of the BHZ model described by Eq. (\ref{eq:BHZ}) with $m=1.2$ and $\gamma=0.2$. (Right) Complex eigen-energy spectrum of the effective Hamiltonian given by Eq. (\ref{eq:BHZ-X}) with $m=1.2$, $\gamma=0.2$, $g_1=2$, and $g_2=1$.}
\label{fig-BHZ}
\end{figure}

The topology of the BHZ model can also be inferred from the edge modes when the system has open boundary condition. When the BHZ model is placed on a cylindrical lattice with open boundary along the $x$ direction, its energy eigenvalues as a function of $k_y$ are shown in the left panel of Figure \ref{fig-BHZ}. Here we assume $m=1.2$ and $\gamma=0.2$ in the topological regime. One can clear see the edge modes corresponding to the non-trivial Fu-Kane invariant.

When the BHZ model is connected to a reservoir, we use the Lindblad equation to describe the open quantum system and first consider the following Lindblad operators
\be
L_{n,i,\uparrow}=\gamma_1a_{n,i,\uparrow}+\gamma_2a_{n+\hat{x},i,\uparrow}\quad \textrm{for}\quad i=1,2.
\ee
Since we only impose Lindblad operators for the spin-up bosons, it clearly breaks the time-reversal invariance. The effective Hamiltonian from the Lindblad equation thus belongs to the A class and describes a non-Hermitian Chern insulator. Transforming to momentum space, the effective Hamiltonian with periodic boundary condition is given by
\be
X=\left(\begin{array}{cc}
         -H_{BHZ}^*+i M/2 & 0 \\
         0 & H_{BHZ}+i M^*/2
\end{array}\right).
\label{eq:BHZ-X}
\ee
where $M$ is given by
\be
M=2\left(\begin{array}{cc}
         g_1+g_2\cos k_x & 0 \\
         0 & -(g_1+g_2\cos k_x)
\end{array}\right).
\ee
Here $g_{1,2}$ depend on $\gamma_{1,2}$, but it is more convenient to treat them as tunable parameters. We have also subtracted a constant matrix in order to make $M$ more symmetric.

The appearance of the $iM/2$ term provides an imaginary part for the eigen-energy. We plot the complex eigen-energy spectrum of the effective Hamiltonian with a cylindrical geometry in the right panel of Figure \ref{fig-BHZ}. In the original BHZ model, the bands of spin-up and spin-down bosons are nearly degenerate. With the Lindblad operators, the bands of the spin-up and spin-down bosons are separated by a finite imaginary part. The separation makes it possible to compute the Chern number for each band even when the eigenvalues are complex numbers. Our numerical calculations show that the two bands with positive real parts have $C=1$ while the other two bands have $C=-1$. In the right panel of Figure \ref{fig-BHZ}, one can see that there exist edge modes between those bands for the BHZ model in a cylinder geometry. Therefore, the Lindblad operators turn the open-system BHZ model into an ordinary Chern insulator by breaking the time-reversal symmetry, characterized by a different topological index when compared to the isolated BHZ model.

\section{Conclusion}\label{sec:conclusion}
Depending on the different settings or quantities used in the analyses, quadratic bosonic systems may fall into different topological classes with different interpretations. The non-Hermitian dynamic matrix is from the bosonic commutation relations and equation of motion, but there is no such complications in fermionic systems with the anticommutation relations. Although the classification tables according to the dynamic matrix of closed system and the effective Hamiltonian from the Lindblad equation of open system are the same, the Lindblad operators may change the symmetry properties and cause a model to be assigned to a different class compared to its closed-system counterpart. 
Our examples have demonstrated the subtle differences in topological properties according to the different settings and spin statistics. Realizations of the models in physical systems or quantum simulators will illustrate the rich physics of topological bosonic systems.

\begin{acknowledgments}
We thank Vincent Flynn for comments. Y. H. was supported by the Natural Science Foundation of China under Grant No. 11874272 and Science Specialty Program of Sichuan University under Grant No. 2020SCUNL210. C. C. C. was supported by the National Science Foundation under Grant No. PHY-2011360.
\end{acknowledgments}

\appendix
\section{Some details}

\subsection{Dynamic matrix}
If we try to diagonalize the Hermitian BdG Hamiltonian of bosons shown in Eq.~(\ref{BdG}) as
\be
V^\dag H V=\Lambda
\label{dia}
\ee
with a diagonal matrix $\Lambda$, then the matrix $V$ must satisfy
\be
V^\dag \tau_3 V=\tau_3
\ee
in order to preserve the bosonic canonical commutation relations. The matrix $V$ is actually a para-unitary matrix. We can rewrite Eq.~(\ref{dia}) as
\be
V^{-1}\Big(\tau_3 H\Big) V=\tau_3\Lambda.
\ee
The dynamic matrix then has the standard structure of the matrix diagonalization. Here the $\tau_3$ factor compensates for the signature in the definition of the para-unitary matrix.

\subsection{Third quantization of bosons}\label{App:3rdQ}
Following the method of Ref.~\cite{Prosen10}, also known as the ``third quantization", one can define a vector space $\mathcal{K}$ that contains the trace class operators and a vector space $\mathcal{K'}$ that contains the unbounded operators, such as physical observables. The elements in $\mathcal{K}$ can be written as a ket $\ket{\rho}$ while the element in $\mathcal{K'}$ can be written as a bra $\bra{A}$. The inner product of these spaces is defined as $\ep{A|\rho}=\textrm{tr}(A\rho)$. For a boson operator $a$, we can define the following left- and right-multiplication maps in $\mathcal{K}$ as follows.
\be
a^L\ket{\rho}=\ket{a\rho},\qquad a^R\ket{\rho}=\ket{\rho a}.
\ee
According to the cyclic property of trace, these maps apply to $\mathcal{K'}$ as follows.
\be
\bra{\rho}a^L=\bra{\rho a},\qquad \bra{\rho}a^R=\bra{a\rho}.
\ee
In terms the above maps of $\mathcal{K}$, we can introduce $4n$ maps in the operator space as $c_{\nu,j}$ and $c'_{\nu,j}$ for $\nu=0,1$ and $j=1,\cdots,n$ as follows.
\be
&&c_{0,j}=a^L_j,~ c'_{0,j}=a^{\dag L}_j-a^{\dag R}_j,~
c_{1,j}=a^{\dag R}_j,~ c'_{1,j}=a^{R}_j-a^{L}_j. \nonumber \\
\ee
The definition of $c'_{\mu,j}$ is designed to right-annihilate the identity operator, $\bra{1}c'_{\mu,j}=0$.
One can verify that they satisfy the canonical commutation relations
\be
[c_{\mu,i},\,c'_{\nu,j}]=\delta_{\mu\nu}\delta_{ij},~
[c_{\mu,i},\,c_{\nu,j}]=0,~
[c'_{\mu,i},\,c'_{\nu,j}]=0.\nonumber\\
\ee
The super-operator $\mathcal{L}$ in the Lindblad equation~\eqref{Lindblad} may be rewritten as
\be
\mathcal{L}=\mathcal{H}^L-\mathcal{H}^R+i\sum_\mu\Big(2L^L_{\mu}L^{\dag R}_{\mu}
-L^{\dag L}_{\mu}L^{L}_{\mu}-L^R_{\mu}L^{\dag R}_{\mu}\Big).\nonumber\\
\label{Liou}
\ee
We want to express the above equation in terms of $c$ and $c'$. To this ends, we reverse the definitions of $c$ and $c'$ and find
\be
a^L_j=c_{0,j},\quad a^{\dag L}_j=c'_{0,j}+c_{1,j}, \nonumber\\
a^R_j=c_{0,j}+c'_{1,j},\quad a^{\dag R}_j=c_{1,j}.
\ee
Together with Eq.~(\ref{eq-H}), it can be shown that
\be
&&\mathcal{H}^L=\sum_{ij}\Big[2A_{ij}(c'_{0,i}+c_{1,i})c_{0,j}+B_{ij}c_{0,i}c_{0,j}\nonumber\\
&&\quad+B^*_{ij}(c'_{0,i}+c_{1,i})(c'_{0,j}+c_{1,j})\Big],\\
&&\mathcal{H}^R=\sum_{ij}\Big[2A_{ij}c_{1,i}(c_{0,j}+c'_{1,j})+B^*_{ij}c_{1,i}c_{1,j}\nonumber\\
&&\quad+B_{ij}(c_{0,i}+c'_{1,i})(c_{0,j}+c'_{1,j})\Big],
\ee
and
\be
&&L_{\mu}^L=\sum_j\Big[l_{\mu j}c_{0,j}+k_{\mu,j}(c'_{0,j}+c_{1,j})\Big],\\
&&L_{\mu}^{\dag L}=\sum_j\Big[l^*_{\mu j}(c'_{0,j}+c_{1,j})+k^*_{\mu,j}c_{0,j}\Big],\\
&&L_{\mu}^R=\sum_j\Big[l_{\mu j}(c_{0,j}+c'_{1,j})+k_{\mu,j}c_{1,j}\Big],\\
&&L_{\mu}^{\dag R}=\sum_j\Big[l^*_{\mu j}c_{1,j}+k^*_{\mu,j}(c_{0,j}+c'_{1,j})\Big].
\ee
Applying the results to Eq.~(\ref{Liou}), the super-operator $\mathcal{L}$ is now in a quadratic form of $c$ and $c'$.
After some algebra, the final result can be found in Eq. (14) of Ref. \cite{Prosen10}. Written in a matrix form, we arrived at Eq.~(\ref{L-op}).


\bibliographystyle{apsrev}

\end{document}